\documentclass{ws-procs10x7}
\usepackage{balance}

\newcolumntype{d}[1]{D{.}{.}{#1}}

\def\Journal#1#2#3#4{{\it #1} {\bf #2}, #3 (#4)}

\newcommand{\dspm}       {\mbox{$D^{\ast \pm}$}}

\newcommand{\dsone}      {\mbox{$D_{s 1}^{+}$}}
\newcommand{\dsonepm}    {\mbox{$D_{s 1}^{\pm}$}}
\newcommand{\done}       {\mbox{$D_1^{0}$}}
\newcommand{\dtwo}       {\mbox{$D_2^{\ast 0}$}}
\makeindex
\begin{document}

\title{CHARM SPECTROSCOPY and EXOTICS AT ZEUS}

\author{L. K. Gladilin$^*$ (On behalf of the ZEUS Collaboration)}

\address{Skobeltsyn Institute of Nuclear Physics\\
Moscow State University,
RU-119992, Vorob'evy Gory, Russia\\
$^*$E-mail: gladilin@sinp.msu.ru }


\twocolumn[\maketitle\abstract{
Light and charmed hadrons are produced copiously in $ep$
collisions with a centre-of-mass energy of $318\,$GeV at HERA.
Results of the ZEUS Collaboration
on pentaquark searches,
deuteron and antideuteron production
and charmed-meson spectroscopy,
obtained using the HERA I data, are summarised.}
]

\section{Introduction}
Light and charmed mesons and baryons are produced copiously in $ep$
collisions with a centre-of-mass energy of $318\,$GeV at HERA.
During the first phase of the HERA operation (1992-2000),
the ZEUS Collaboration accumulated a data sample corresponding
to an integrated luminosity of $\sim120\,$pb$^{-1}$.
Results
on pentaquark searches,
deuteron and antideuteron production
and charmed-meson spectroscopy,
obtained by the ZEUS Collaboration
using the HERA I data, are summarised
in this note.

\section{Strange pentaquarks}
A peak in the $K^0_s p ({\bar{p}})$ invariant mass spectrum around $1520\,$MeV
was observed in deep inelastic scattering
(DIS) by the ZEUS Collaboration~\cite{zeus_theta}.
In Fig.~1, the spectrum is shown for
$Q^2>20\,$GeV$^2$,
where $Q^2$ is
the exchanged-photon virtuality
The statistical significance of the signal varies between 3.9$\,\sigma$
and 4.6$\,\sigma$ depending upon the treatment of the background.
The candidate $\Theta^+$ signal was found to be produced predominantly
in the forward (proton) direction in the laboratory
frame~\cite{zeus_theta_study}.
This is unlike the case for the $\Lambda(1520)$ and can indicate that
$\Theta^+$ may have an unusual production
mechanism related to the proton-remnant fragmentation.
Using a smaller integrated luminosity,
the H1 Collaboration did not observe
the candidate $\Theta^+$ signal
and set an upper limit
on its production~\cite{h1_theta}.
The production cross section for the $\Theta^+$ candidate, measured by the
ZEUS Collaboration~\cite{zeus_thexs}, is
$125\pm27({\rm stat.})^{+37}_{-28}({\rm syst.})\,{\rm pb}$
for $Q^2>20\,$GeV$^2$. The value is larger than, but consistent with, the upper
limit obtained by the H1 Collaboration.

\begin{figure}[b]
\centerline{\psfig{file=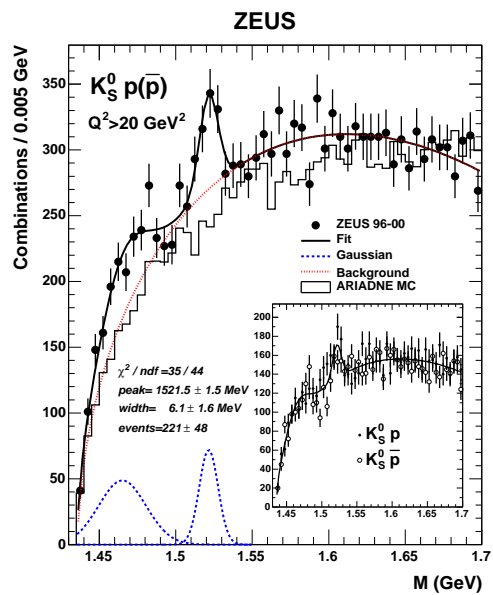,width=2.6in}}
\caption{
Invariant mass spectrum for the $K^0_s p ({\bar{p}})$ combinations
obtained by the ZEUS Collaboration.}
\label{fig1}
\end{figure}

The ZEUS Collaboration performed also a search
for two strange pentaquarks, reported by
the NA49 Collaboration~\cite{na49_pq},
and observed no signal in the $\Xi \pi$ invariant mass spectrum~\cite{zeus_xi}.
Upper limits on the ratio of a possible $\Xi_{3/2}^{--}(\Xi_{3/2}^0)$
signal to the $\Xi^0(1530)$ signal were set in the mass range
$1.65-2.35\,$GeV.

\section{Deuterons and antideuterons}

A first observation of deuteron and antideuteron production
in $ep$ collisions in the DIS regime has been recently reported by the ZEUS
collaboration~\cite{zeus_deuteron}.
To identify the particles,
the measurement of
the energy losses, $dE/dx$, in the central tracking detector
has been used.
\begin{figure}[b]
\centerline{\psfig{file=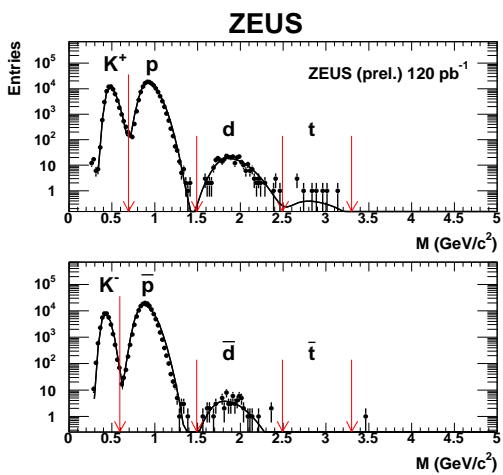,width=2.7in}}
\caption{
The mass spectra for positive and negative tracks 
with $dE/dx>2.5\,$mips.
The arrows show cuts applied for selection of (anti)proton and
(anti)deuteron candidates.}
\label{fig2}
\end{figure}

Figure~2 shows the mass spectra for positive and negative tracks
with $dE/dx>2.5\,$mips (minimum ionising particles)
in DIS events with $Q^2>1\,$GeV$^2$.
For each track, the mass has been calculated from the track momentum
and $dE/dx$ value
using a parametrisation based on the Bethe-Bloch equation.
The number of deuteron and antideuteron candidates in the mass window
$1.5<M<2.5\,$GeV is 309 and 62, respectively. No antitritons
have been observed, Given the small number of the triton candidates,
no statement on the triton observation in DIS has been made.

Small contributions to the observed signals from the secondary interactions
in the beam-pipe and detector material have been subtracted using side bands of
the distribution on the distance of the closest approach of a track to
the beam spot in the transverse plane.
Possible contamination of the observed deuteron signal
due to interactions of the proton or electron beam with the residual
gas in the beam-pipe has been found to be below $20\%$.
\begin{figure}[b]
\centerline{\psfig{file=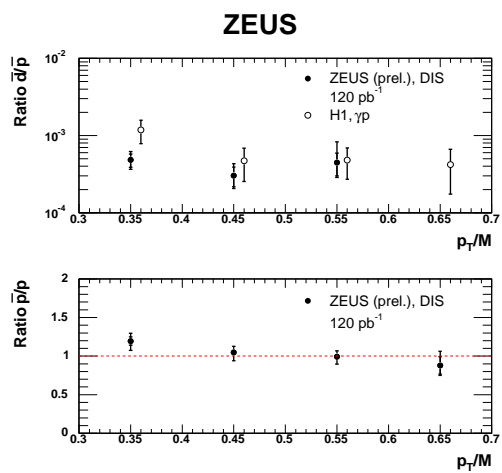,width=2.7in}}
\caption{
The corrected (top) ${\bar d}/{\bar p}$ and
(bottom ) ${\bar p}/p$ production ratios as a function
of $p_T/M$.
Open circles show the ${\bar d}/{\bar p}$ ratios obtained
by the H1 Collaboration in the photoproduction regime.}
\label{fig3}
\end{figure}

Figure~3 shows the ${\bar d}/{\bar p}$ and
${\bar p}/p$ production ratios, corrected for the detector and trigger effects,
as a function of $p_T/M$.
The measured ${\bar p}/p$ ratio is consistent with unity.
The ${\bar d}/{\bar p}$ ratios are in fair agreement with
those obtained in $pp$ interactions~\cite{pp_deuteron},
in hadronic $\Upsilon$ decays~\cite{argus_deuteron},
and by the H1 Collaboration at HERA in
the photoproduction regime ($Q^2<1\,$GeV$^2$)~\cite{h1_deuteron}.

No antideuteron candidates have been found in the current region of
the Breit frame~\cite{breit}. Since the region is similar to a single
hemisphere of the $e^+e^-$ annihilation process, the antideuteron observation
in DIS does not contradict to the small antideuteron rates observed
at LEP~\cite{lep_deuteron}.

\section{Charmed pentaquark}

An observation of a candidate
for the charmed pentaquark state, $\Theta^0_c = uudd{\bar c}$,
decaying to $D^{*\pm}p^\mp$
was reported by the H1 Collaboration~\cite{h1_ch5q}.
A fit of the signal in DIS yielded $50.6\pm11.2$ signal events and
the mass of $3099\pm3({\rm stat.})\pm5({\rm syst.})\,$MeV.
The observed resonance was reported to contribute
$(1.46\pm0.32)\%$ of the $D^{*\pm}$ production rate
in the kinematic range studied in DIS~\cite{h1_thetac_study}.
The measured differential distributions were found to be
consistent with the fragmentation modelling of the candidate $\Theta^0_c$
production.

\begin{figure}[b]
\centerline{\psfig{file=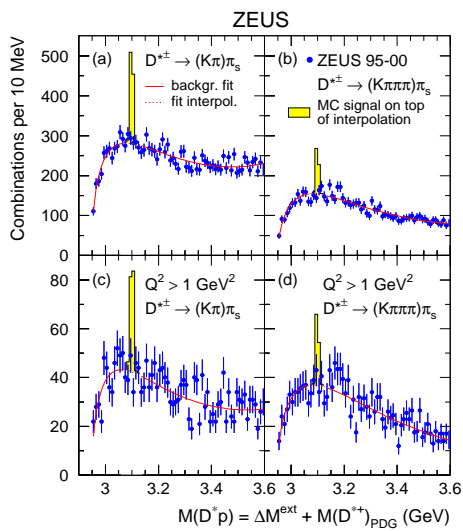,width=2.7in}}
\caption{
The distributions of $M(D^{*\pm}p^\mp)$
obtained by the ZEUS Collaborations.}
\label{fig4}
\end{figure}

The observation of the H1 Collaboration was challenged
by the ZEUS Collaboration~\cite{zeus_ch5q}.
Using a larger sample of $D^{*\pm}$ mesons,
ZEUS observed no signature of the narrow resonance in
the $M(D^{*\pm}p^\mp)$ spectra shown in Fig.~4.
The Monte Carlo $\Theta^0_c$ signals normalised to $1\%$
of the number of reconstructed $D^{*\pm}$ mesons are shown
on top of the fitted backgrounds.
The upper limit on the fraction of $D^{*\pm}$ mesons
originating from $\Theta^0_c$ decays was evaluated to be 
$0.23\%$ ($95\%$ C.L.). The upper limit for DIS with $Q^2>1\,$GeV$^2$
is $0.35\%$ ($95\%$ C.L.).

\section{Studies of excited $D$ mesons}

Sizeable
production of the excited charmed, $D^0_1$ and $D_2^{*0}$,
and charmed-strange, $D_{s1}^+$,
mesons has
been observed in $ep$ interactions
by the ZEUS collaboration~\cite{zeus_d1d2,zeus_d1s}.
Figure~5 shows the distribution
of the $M(\dspm K^0_s)$
for the $\dspm$ candidates.
A clear signal is seen
at the nominal value of $M(\dsone )$.
The measured $\done$, $\dtwo$ and $\dsone$ rates were converted to
the fractions of $c$ quarks hadronising as a particular excited charmed meson.
These fragmentation fractions agree with previous measurements
in $e^+e^-$ annihilations.

\begin{figure}[b]
\centerline{\psfig{file=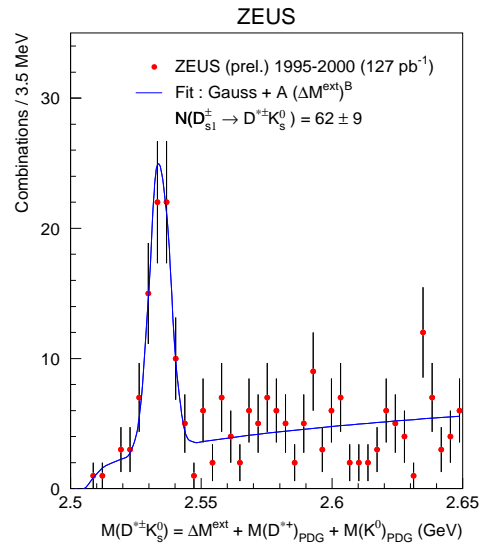,width=2.8in}}
\caption{
The distributions of the
$M(\dspm K^0_s)$
for the $\dspm$ candidates.}
\label{fig5}
\end{figure}

The helicity angular distribution of the \dsonepm\ signal
was fitted
to a form $1+R\cos^2\alpha$, where $R$ is a helicity
parameter and
the helicity angle ($\alpha$) is defined as the angle between
the $K_S^0$ and $\pi_S$ momenta in the \dspm\ rest frame.
The unbinned likelihood fit yielded the value
$R=-0.53\pm0.32({\rm stat.})^{+0.05}_{-0.14}({\rm syst.})$.
This value is consistent with the helicity parameter value
$-0.23^{+0.40}_{-0.32}$ obtained
by the CLEO Collaboration for $\dsonepm$ meson
in the $D^{*0}K^{\pm}$ final state~\cite{cleo_d1s}.
The present measurement does not contradict the conclusion
of the CLEO Collaboration that $R=0$ and, thus, the spin-parity of
the \dsone\ is $1^+$. However, this measurement is also consistent
with $R=-1$, expected for spin-parities
$1^-$ and $2^+$~\cite{godfrey_kokoski}.
Recently, a precise measurements of the $\dsone$ helicity parameter,
$R=-0.70\pm0.03$, has been reported by the Belle
Collaboration~\cite{belle_d1s}.
The measured value has been interpreted as a result of the $D$ and $S$
waves mixture due to an interference of the $\dsone$ meson
with the recently discovered $D_{sJ}(2460)^+$ meson.

The ZEUS search for the radially excited $D^{*\prime\pm}$ meson,
reported by the DELPHI Collaboration~\cite{delphi_drad},
revealed
no signal~\cite{zeus_d1d2}.
The upper limit on
the product of the fraction of $c$ quarks hadronising as
a $D^{*\prime +}$ meson
and the branching ratio of the $D^{*\prime +}$ decay to
$D^{*+}\pi^+\pi^-$
was estimated to be $0.7\%$ ($95\%$ C.L.).
This limit is somewhat stronger than
the $0.9\%$ limit obtained by the OPAL Collaboration~\cite{opal_drad}.

\balance

\end{document}